\documentclass[prb,twocolumn,showpacs,preprintnumbers,amsmath,amssymb,superscriptaddress,floatfix]{revtex4}
\usepackage{graphicx,subfigure}
\usepackage{dcolumn}
\usepackage{bm}
\usepackage{psfrag}
\usepackage{multirow}

\def\be{\begin{equation}}       \def\ee{\end{equation}}
\def\bea{\begin{eqnarray}}      \def\eea{\end{eqnarray}}
\def\half{\frac{1}{2}}
\def\dag{\dagger}
\def\non{\nonumber}
\begin{document}
\title{Non-linear spin wave theory results for the  
frustrated $S=\half$ Heisenberg antiferromagnet on a body-centered cubic lattice}
\author{Kingshuk Majumdar}
\affiliation{Department of Physics, Grand Valley State University, Allendale, 
Michigan 49401, USA}
\email{majumdak@gvsu.edu}
\author{Trinanjan Datta}
\affiliation{Department of Chemistry and Physics, Augusta State University, Augusta, 
Georgia 30904, USA}
\email{tdatta@aug.edu}
\date{\today}

\begin{abstract}\label{abstract}
At zero temperature the sublattice magnetization of the quantum spin-$1/2$ Heisenberg antiferromagnet on a body-centered cubic lattice with competing first and second neighbor exchange ($J_1$ and $J_2$) is 
investigated using the non-linear spin wave theory. The zero temperature phases of the model consist of a two sublattice N\'{e}el phase for small $J_2$ (AF$_1$) and a collinear phase 
at large $J_2$ (AF$_2$). We show that quartic corrections due to spin-wave interactions enhance the sublattice magnetization in both the AF$_1$ and the AF$_2$ phase. The magnetization corrections are prominent near the classical transition point of the model and in the $J_2> J_1$ regime. The ground state energy with quartic interactions is also calculated. It is found that up to quartic corrections the first order phase transition (previously observed in this model) between the AF$_1$ and 
the AF$_2$ phase survives. 
\end{abstract}
\maketitle

\section{\label{sec:Intro}Introduction}
In recent years thermodynamic properties of frustrated quantum Heisenberg
antiferromagnets have been of intense interest both theoretically 
and experimentally in condensed matter physics.~\cite{diep,subir1} 
The phase diagram of the quantum spin-$1/2$ Heisenberg antiferromagnetic (AF)
model on two-dimensional (2D) lattices with nearest neighbor ($J_1$) and next nearest neighbor interactions ($J_2$) have been studied extensively by different methods.~\cite{tassi1,tassi2,
subir2,subir3,harris,dot,huse,chubu1,chubu2,gelfand,sushkov,weihong,
singh,oleg,valeri,an,honda,svistov,gochev} For the square lattice with nearest neighbor 
(NN) exchange interaction only, the ground state is antiferromagnetically
ordered at zero temperature. Addition of next nearest neighbor (NNN) interactions 
break the AF order. The competition between the NN and NNN interactions
for the square lattice is characterized by the frustration parameter $p$. 
It has been found that a quantum spin liquid phase exists between $p_{1c} \approx 0.38$ 
and $p_{2c} \approx 0.60$.
For $p<p_{1c}$ the lattice is AF-ordered whereas for $p>p_{2c}$ a collinear phase
emerges. In the collinear state the NN spins have a parallel orientation in the vertical
direction and antiparallel orientation in the horizontal direction or vice versa.

Motivated by the results for the 2D lattices some work has been done by analytical
and numerical techniques to understand the magnetic phase diagram of three dimensional (3D) 
lattices.~\cite{oguchi,katanin,oitmaa,Schmidt,viana} Linear spin-wave theory, 
exact diagonalization, and linked-cluster series expansions (both at zero and finite 
temperature) have been utilized to study the 3D quantum spin-$1/2$ Heisenberg AF on a 
body-centered-cubic lattice (bcc) lattice.~\cite{Schmidt,oitmaa} It has been found 
that the lattice does not have a quantum disordered phase and a first-order 
phase transition from the AF-phase (AF$_1$) to lamellar state (AF$_2$) occurs 
at $p_c=0.53$ or J$_2$/J$_1 \approx$ 0.705. The first-order nature of the phase 
transition from the AF$_1$ to the AF$_2$ phase in the model is inferred from a kink 
in the ground state energy of the system. In one and 2D due to reduced phase space 
quantum fluctuations play an important role in determining the quantum critical 
points of the system at low temperature. However, in 3D the phase space available 
is greater and quantum fluctuations play a lesser role. Hence, the absence of the 
quantum disordered phase for the BCC lattice. 

In this work, we study the 3D quantum spin-$1/2$ AF on a bcc lattice using the 
non-linear spin wave theory where we consider interactions between spin waves up to 
quartic terms in the Hamiltonian. We compute the effect of these higher order terms 
on the sublattice magnetization (see Fig~\ref{fig:submag}). The corrections to the 
magnetization become important as the classical transition point p$_c$=0.5 is 
approached. Also, our calculations re-confirm the first order nature of the phase 
transition found in Refs.~\onlinecite{Schmidt,oitmaa} up to quartic interactions 
(see Fig~\ref{fig:Energy}).
\begin{figure}[httb]
\centering
\subfigure[\;AF$_1$ phase]
{\includegraphics[width=2.5in]{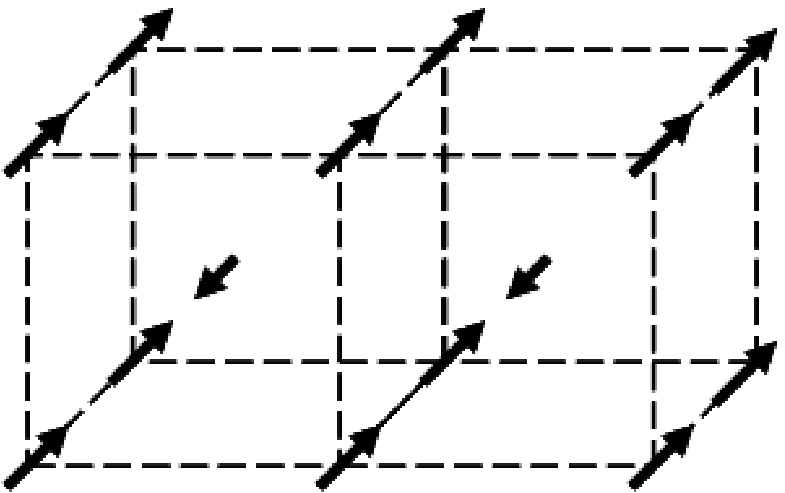}}
\hfill
\subfigure[\;AF$_2$ phase]{
\includegraphics[width=2.5in]{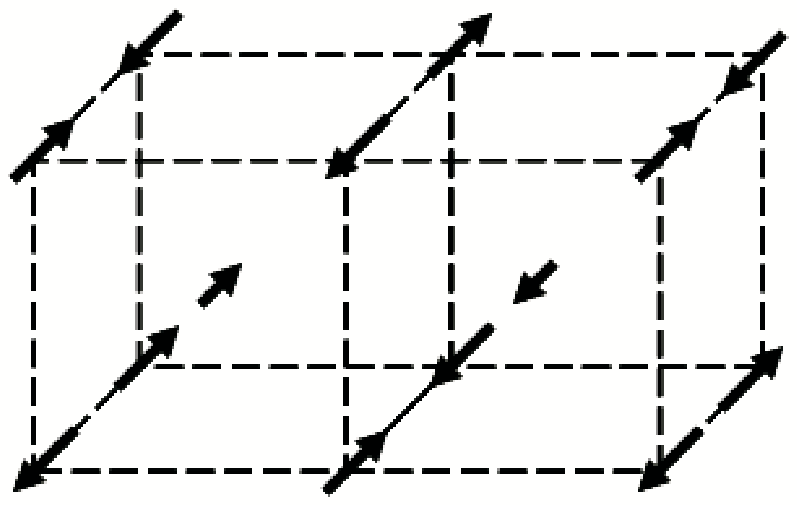}}
\caption{\label{fig:AF} AF$_1$ and AF$_2$ ordered phases of the bcc lattice.
In the AF$_1$ phase all $A$-sublattice spins point in the direction of an 
arbitrary unit vector while $B$-sublattice spins point in the 
opposite direction. For the AF$_2$ phase there are two interpenetrating N\'{e}el 
states each living on the initial sublattices $A$ and $B$.}
\end{figure}
The paper is organized as follows. In Section~\ref{sec:formalism} we begin 
with a brief description of the properties of the bcc lattice
relevant to our calculations. We then set-up the Hamiltonian 
for the Heisenberg spin-$1/2$ AF on the bcc lattice. 
The classical ground state
configurations of the model and the different phases are then discussed. 
Next we map the spin Hamiltonian to the Hamiltonian of interacting bosons and 
the non-linear spin-wave theory for the two phases are developed. The sublattice 
magnetizations and the ground state energies for the two phases are numerically 
calculated and the results are plotted and discussed in Section~\ref{sec:results}. 
Finally we summarize our results in Section~\ref{sec:conclusions}.

\section{\label{sec:formalism} Formalism}
Body-centered-cubic lattice consists of two interpenetrating, identical simple
cubic lattices, each of which consists of two interpenetrating, identical 
face-centered lattices. This makes the bcc lattice a 3D bi-bipartite 
cubic lattice. The basis vectors of the bcc lattice connecting eight ($z_1=8$) 
nearest neighbors are (in units of simple cubic lattice spacing)
${\bf a_1}=(1,1,-1),\; {\bf a_2}=(1,-1,1),\;{\bf a_3}=(-1,1,1)$ and the lattice 
vectors connecting six ($z_2=6$) next-nearest neighbors are 
${\bf b_1}=(\pm 2,0,0),\; {\bf b_2}=(0,\pm 2,0)$ and ${\bf b_3}=(0,0,\pm 2)$. On such a 
lattice the Hamiltonian for a spin-$1/2$ Heisenberg AF with first and second 
neighbor interactions is 
\be
H = \half J_1 \sum_{\langle ij \rangle}{\bf S}_i \cdot {\bf S}_j 
+ \half J_2 \sum_{[ij]}{\bf S}_i \cdot {\bf S}_j.
\label{ham}
\ee
where $J_1$ is the NN and $J_2$ is the frustrating NNN exchange constants. Both 
couplings are considered AF, i.e. $J_1,J_2>0$. 

\subsection{\label{sec: classical}Classical ground state configurations}

The limit of infinite spin, $S \rightarrow \infty$ corresponds 
to the classical Heisenberg model. We assume that the set of possible
spin configurations of the system are described by 
$S_i=S{\bf u}e^{i{\bf q}\cdot{\bf r_i}}$,
where ${\bf u}$ is a vector expressed in terms of an arbitrary orthonormal basis and
${\bf q}$ defines the relative orientation of the spins on the lattice.~\cite{villain} 
The classical ground state energy of the system expressed as a function of the 
parameters $J_1$ and $J_2$ takes the form
\be
E_{\bf k}/NJ_1
= \half S^2z_1[\gamma_{1 {\bf k}}+p\gamma_{2 {\bf k}}], 
\ee
with the structure factors
\begin{eqnarray}
\gamma_{1{\bf k}} &=&\cos(k_x)\cos(k_y)\cos(k_z), \\
\gamma_{2{\bf k}} &=& \Big[\cos(2k_x)+\cos(2k_y)+\cos(2k_z)\Big]/3.
\end{eqnarray}
where $N$ is the number of sites on the lattice and 
$p=z_2J_2/z_1J_1$ is defined to be the parameter of frustration.

At zero temperature, the classical ground state for the bcc lattice has 
two phases. In the limit of small $p$ or $J_2 << J_1$ three isolated 
minima in energy, $E_0/NJ_1=-4S^2(1-p)$ occur at the wave-vectors
$(\pi,0,0), (0,\pi,0)$ and $(0,0,\pi)$. They correspond to the 
classical two-sublattice N\'{e}el state
(AF$_1$ phase) where all $A$-sublattice spins point in the direction of an 
arbitrary unit vector ${\bf {\hat n}}$ while $B$-sublattice spins point in the 
opposite direction $-{\bf {\hat n}}$. 

In the other limit, for large $p$ or $J_2 >> J_1$, there is 
a single minimum in energy, $E_0/NJ_1=-4S^2p$ 
at ${\bf k}=(\pi/2,\pi/2,\pi/2)$. In this case the classical ground state 
consists of two interpenetrating N\'{e}el states (AF$_2$ phase)
each living on the initial sublattices $A$ and $B$. The two phases are shown in
Fig.~\ref{fig:AF}.

The classical limit for the phase transition from 
AF$_1$ to AF$_2$ for the 3D model on the bcc lattice is at
the critical value $p_c=2z_2/3z_1=1/2$ i.e. when $J_2/J_1=2/3$. 
This is similar to the spin-$1/2$ $J_1-J_2$ model on a 2D square 
lattice where the critical value of $p_c=1/2$ or $J_2/J_1=1/2$.

\subsection{\label {spinwave} Non-linear spin wave theory}
The Hamiltonian in Eq.~\ref{ham} can be mapped into an equivalent Hamiltonian
of interacting bosons by transforming the spin operators to bosonic operators 
$a,a^\dag$ for $A$ sublattice and $b, b^\dag$ for $B$ sublattice using the 
well-known Holstein-Primakoff transformations~\cite{holstein}
\begin{eqnarray}
S_{Ai}^+ &\approx& \sqrt{2S}\Big(1- \frac {a_i^\dag a_i}{4S} \Big)a_i,\;\;
S_{Ai}^- \approx\sqrt{2S}a_i^\dag \Big(1-\frac {a_i^\dag a_i}{4S}  \Big), \non \\
S_{Ai}^z &=& S-a^\dag_ia_i,  \non \\ 
S_{Bj}^+ &\approx&\sqrt{2S}b_j^\dag \Big(1-\frac {b_j^\dag b_j}{4S}\Big),\;\;
S_{Bj}^- \approx\sqrt{2S}\Big(1-\frac {b_j^\dag b_j}{4S}  \Big)b_j, \non \\
S_{Bj}^z &=& -S+b^\dag_jb_j,\label{holstein}
\end{eqnarray}
In these transformations we have kept terms up to the order of $1/S$. Next using the
Fourier transforms
\[
a_i = \sqrt{\frac 2{N}}\sum_{\bf k} e^{-i{\bf k \cdot R_i}}a_{\bf k},\;\;\;
b_j = \sqrt{\frac 2{N}}\sum_{\bf k} e^{-i{\bf k \cdot R_j}}b_{\bf k},
\]
the real space Hamiltonian is transformed to the ${\bf k}$-space Hamiltonian.
The reduced Brillouin zone contains $N/2$ ${\bf k}$ vectors as the unit cell is 
a magnetic supercell consisting of an $A$-site and a $B$-site. In the following two
sections we study the cases $J_2<J_1$ and $J_2>J_1$ separately.

\subsubsection{\label{sec: smallp}$J_2<J_1$: AF$_1$ phase}
In this phase the classical ground state is the two-sublattice N\'{e}el state 
(see Fig.~\ref{fig:AF}). For
the NN interaction spins in $A$ sublattice interacts with spins in
$B$ sublattice and vice-versa. On the other hand for the NNN 
exchange $J_2$ connects spins on the same sublattice $A$ with $A$ and $B$ with
$B$. Substituting Eq.~\ref{holstein} in Eq.~\ref{ham}, 
expanding the radical, and restricting to terms only up to the anharmonic 
quartic terms, we obtain the ${\bf k}$-space Hamiltonian 
\be
H=H^{(0)}+H^{(2)}+ H^{(4)}.
\label{smallpH}
\ee
The classical ground state energy $H^{(0)}$ and the 
quadratic terms $H^{(2)}$ are
\bea
H^{(0)} &=& -\half NJ_1S^2 z_1 (1-p)
\label{cgs}\\
H^{(2)} &=& J_1S z_1 \sum_{\bf k}\Big[A_{0{\bf k}}(a_{\bf k}^\dag a_{\bf k}
 +b_{\bf k}^\dag b_{\bf k}) 
\nonumber \\&+&
B_{0{\bf k}}(a^\dag_{\bf k}b^\dag_{-\bf k}+a_{-\bf k}b_{\bf k})\Big],
\label{smallJ2}
\eea
with the coefficients $A_{0{\bf k}}$ and $B_{0{\bf k}}$ defined as
\begin{eqnarray}
A_{0{\bf k}}&=&1-p(1-\gamma_{2{\bf k}}), \label{a0smallp}\\ 
B_{0{\bf k}}&=& \gamma_{1{\bf k}}.
\label{b0smallp}
\end{eqnarray}
The quartic terms in the Hamiltonian $H^{(4)}$ are
\bea
H^{(4)} &=&-J_1 \sum_{\langle ij \rangle}\Big[a_i^\dag a_i b^\dag_j b_j 
+\frac 1{4}\Big(a_ib^\dag_jb_jb_j+a_i^\dag a_ia_ib_j 
+ h.c.\Big)\Big] \non \\
&+&\half J_2\sum_{[ij]}\Big[a^\dag_ia_ia^\dag_ja_j-\frac 1{4}\Big(a_ia^\dag_ja_j^\dag a_j 
+ a_i^\dag a_ia_ia_j^\dag+h.c.\Big)\nonumber\\&+& a \leftrightarrow b\Big]. 
\label{quartic}
\eea
These terms are evaluated by applying the Hartree-Fock
decoupling process.~\cite{cshev} In the harmonic approximation the following
Hartree-Fock averages are non-zero for the bcc-lattice Heisenberg 
AF:
\bea
u &=& \langle a_i^\dag a_i \rangle = \langle b_i^\dag b_i \rangle = 
\half \Big[\frac 2{N} \sum_{\bf k}
\frac {A_{0{\bf k}}}{\omega_{0{\bf k}}} -1\Big],  \label{ucoeff}\\
v &=& \langle a_i b_j \rangle = \langle a_i^\dag b_j^\dag \rangle
=-\half \Big[\frac 2{N} \sum_{\bf k}
\frac {\gamma_{1{\bf k}}B_{0{\bf k}}}{\omega_{0{\bf k}}}\Big],\label{vcoeff}\\
w &=& \langle a_i^\dag a_j \rangle = \langle b_i^\dag b_j \rangle =
\half \Big[\frac 2{N} \sum_{\bf k}
\frac {\gamma_{2{\bf k}}A_{0{\bf k}}}{\omega_{0{\bf k}}}\Big], 
\label{wcoeff}
\eea 
where $\omega_{0{\bf k}}=\sqrt{A_{0{\bf k}}^2 -B_{0{\bf k}}^2}.$
 
The contributions of the decoupled quartic terms 
to the harmonic Hamiltonian in Eq.~\ref{smallJ2} are to renormalize the values
of $A_{0{\bf k}}$ and $B_{0{\bf k}}$ which are now
\bea
A_{\bf k} &=& \Big( 1-\frac {u+v}{S}\Big) 
- p[1-\gamma_{2 {\bf k}}]\Big( 1- \frac {u-w}{S} \Big),  
\label{Acoeff}\\
B_{\bf k} &=& \gamma_{1{\bf k}}\Big( 1-\frac {u+v}{S} \Big),
\label{Bcoeff}\\
\omega_{\bf k} &=& \sqrt{A_{\bf k}^2 -B_{\bf k}^2}.\label{omega}
\eea 

The quartic corrections to the ground state energy is calculated from the four-boson
averages. In the leading order they are decoupled into the bilinear combinations
(Eqs.~\ref{ucoeff} --~\ref{wcoeff}) using Wick's theorem. The corresponding four boson 
terms are,
\bea
\langle a^\dag_i a_i b^\dag_j b_j \rangle &=& u^2+v^2,\;\;\; 
\langle a^\dag_i b^\dag_j b_jb_j \rangle = 2uv,
 \non \\
\langle a^\dag_i a_i a_i b_j \rangle &=& 2uv,\;\;\;  
\langle a^\dag_i a_i a_j^\dag a_j \rangle = u^2+w^2,  \\
\langle a_i a_j^\dag a_j^\dag a_j \rangle &=& 2uw, \;\;\;
\langle a^\dag_i a_i a_i a_j^\dag \rangle = 2uw. \non 
\eea 
This yields the ground state energy correction from the quartic terms
\be
\delta E^{(4)}= -\half NJ_1z_1\Big[(u+v)^2-p(u-w)^2\Big].
\label{EquarticsmallJ}
\ee
Summing all the corrections together the ground state energy takes the form
\bea
E/NJ_1 &=&-\half z_1S(S+1)(1-p)+
\half z_1S \Big[\frac 2{N} \sum_{\bf k} \omega_{\bf k}\Big] \non \\
&+& \half z_1\Big[(u+v)(1-u-v) - p(u-w)(1-u+w)\Big].\nonumber\\
\label{energy}
\eea
and the sublattice magnetization $\langle S_\alpha \rangle$ at zero temperature 
is given by
\be
\langle S_\alpha \rangle = S\Big[1-\frac 1{2S}
\Big\{\frac 2{N} \sum_{\bf k}\frac {A_{\bf k}}{\omega_{\bf k}} -1\Big\}\Big].
\label{mag}
\ee
Using Eqs.~\ref{Acoeff},~\ref{Bcoeff}, and ~\ref{omega} we numerically evaluate $E/NJ_1$ 
and $\langle S_\alpha \rangle$. For the bcc lattice the {\bf k}-sum is replaced by an 
integral over the Brillouin zone~\cite{flax}
\be
\frac 2{N}\sum_{\bf k} \rightarrow \frac 1{\pi^3}\int_0^\pi \int_0^\pi
\int_0^\pi\; dk_x dk_y dk_z. 
\ee

\subsubsection{\label{sec: large_mu}$J_2>J_1$: AF$_2$ phase}
The classical ground state for $J_2>J_1$ corresponds to a four sublattice state
where each of the $A$ and $B$ sublattice is itself antiferromagnetically 
ordered (see Fig.~\ref{fig:AF}). For the NN exchange there 
are four $A-A$, four $B-B$, and eight $A-B$ type interactions between the 
sublattices. In case of NNN exchanges there are a total of twelve
$A-B$ type interactions. Adding all their contributions together up to the quadratic
terms the harmonic Hamiltonian takes the same form as Eq.~\ref{smallJ2} with
\bea
H^{(0)} &=& -\half NJ_1S^2z_1p, \\
A_{0{\bf k}}&=& \half (\gamma_{1{\bf k}}+2p),\label{A0}\\ 
B_{0{\bf k}}&=& \half (\gamma_{1{\bf k}}+2p\gamma_{2{\bf k}}).
\label{B0}
\eea
The quartic terms in the Hamiltonian for this case are
\bea
H^{(4)} &=&-J_1 \sum_{\langle ij \rangle}\Big[a_i^\dag a_i b^\dag_j b_j 
+\frac 1{4}\Big(a_ib^\dag_jb_jb_j+a_i^\dag a_ia_ib_j 
+ h.c.\Big)\Big] \non \\
&+&\half J_1\sum_{\langle ij \rangle}\Big[a^\dag_ia_ia^\dag_ja_j
-\frac 1{4}\Big(a_ia^\dag_ja_j^\dag a_j + a_i^\dag a_ia_ia_j^\dag 
+ h.c.\Big)\non\\&+& a \leftrightarrow b\Big] -J_2 \sum_{\langle ij \rangle}\Big[a_i^\dag a_i b^\dag_j b_j
+ \frac 1{4}\Big(a_ib^\dag_jb_jb_j+ a_i^\dag a_ia_ib_j \non\\&+& h.c.\Big)\Big] .
\label{quartlargeJ}
\eea 
These terms are decoupled and evaluated in the same way as before. The 
renormalized values of the coefficients $A_{\bf k}$ and $B_{\bf k}$ are 
\bea
A_{\bf k} &=& \half \Big[ \gamma_{1{\bf k}}\Big(1-\frac {u-{\overline w}}{S}
 \Big)-\frac {v+{\overline w}}{S}  
+ 2p\Big(1-\frac {u+{\overline v}}{S} \Big)\Big], 
\label{largeA}\non\\\\
B_{\bf k} &=& \half \Big[\gamma_{1{\bf k}}\Big(1-\frac {u+v}{S} \Big) 
+ 2p\gamma_{2{\bf k}}\Big(1-\frac {u+{\overline v}}{S} \Big)\Big], 
\label{largeB}
\eea 
where
\bea
{\overline v} &=& -\half \Big[\frac 2{N} \sum_{\bf k}
\frac {\gamma_{2{\bf k}}B_{0{\bf k}}}{\omega_{0{\bf k}}}\Big],\\
{\overline w} &=& 
\half \Big[\frac 2{N} \sum_{\bf k}
\frac {\gamma_{1{\bf k}}A_{0{\bf k}}}{\omega_{0{\bf k}}}\Big]. 
\eea
In Eqs.~\ref{largeA} and \ref{largeB} $u,v$ have the same form as in 
Eqs.~\ref{ucoeff} --~\ref{vcoeff} but they
are evaluated with the coefficients $A_{0\bf k}$ and $B_{0\bf k}$ in 
Eqs.~\ref{A0} and ~\ref{B0}. The quartic corrections to the ground state 
energy is
\be
\delta E^{(4)} = \half NJ_1z_1 \Big\{(u-{\overline w})^2 -(u+v)^2 
-p(u+{\overline v})^2 \Big\}.
\ee
Combining all these corrections the ground state energy is
\bea
E/NJ_1 &=& -\half z_1 S(S+1)p 
+\half z_1S \Big[\frac 2{N}\sum_{\bf k} \omega_{\bf k}\Big] \non \\ 
&+& \frac 1{4} z_1 \Big[(v+{\overline w})(1-2u-v+{\overline w}) \non\\
&+& 2p(u+{\overline v})(1-u-{\overline v})\Big].
\label{gshighp}
\eea
The sublattice magnetization and ground state energy are then obtained 
numerically using Eqs.~\ref{mag},~\ref{largeA},~\ref{largeB}, and~\ref{gshighp}. 

\section{\label{sec:results}Results}
\begin{figure}[httb]
\centering
\includegraphics[width=3in]{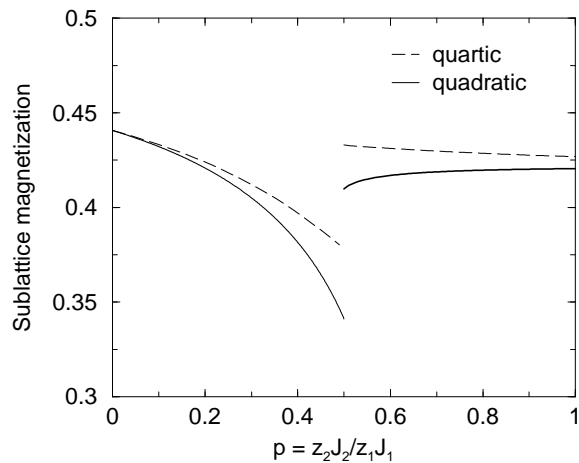}
\caption{\label{fig:submag} Sublattice magnetization, $\langle S_\alpha  \rangle $, 
is plotted versus $p$ for AF$_1$ and AF$_2$ ordered phases. 
In the AF$_1$ phase with increase in
$p$ the system aligns the spins antiferromagnetically along the horizontal and 
the vertical directions -- thus decreasing the sublattice magnetization. 
In the AF$_2$ phase $\langle S_\alpha  \rangle $ mostly stays
the same and then shows a slight decrease (without quartic corrections) as $p$ 
approaches the critical value
$p_c=0.5$ from above. However with the quartic corrections 
$\langle S_\alpha  \rangle $ remains almost constant at $\approx 0.43$. In both cases 
quartic corrections to the Hamiltonian of the system enhance the magnetic order.}
\end{figure}

In Fig.~\ref{fig:submag} we show the results for the sublattice magnetization, 
$\langle S_\alpha \rangle $, obtained numerically from Eq.~\ref{mag} for 
both AF$_1$ and AF$_2$ phases with (dashed line) and without (solid line) quartic 
corrections. In the
AF$_1$ ordered phase or the two sublattice N\'{e}el phase where $A$ and $B$ sublattice
spins point in the opposite directions, sublattice 
magnetization decreases monotonically
with increase in $p$ till $p \approx 0.5$. 
The curve starts at $\approx 0.44$ for $p=0$ and ends at $\approx 0.34$ for $p=0.5$.
The gradual decrease in $\langle S_\alpha \rangle $ is expected with increase in $p$ as 
increasing strength of NNN interaction $J_2$ aligns the spins 
antiferromagnetically along the horizontal and the vertical directions. The quartic 
corrections produce a change in the sublattice magnetization, 
$\langle S_\alpha \rangle$, which becomes significant as one approaches the 
classical transition point p$_c$=0.5 (see Fig.~\ref{fig:submag}). With quartic 
corrections the magnetization curve starts at $\approx 0.44$ for $p=0$ and ends 
at $\approx 0.38$ for $p=0.49$. At $p=0$ (no frustration) there is no quartic corrections to
$\langle S_\alpha \rangle $.
This can be observed from Eqs.~\ref{Acoeff}--~\ref{omega},~\ref{mag} as the 
correction factor $(1-(u+v)/S)$ cancels out in Eq.~\ref{mag}. 
At the wave-vector ${\bf k}=(\pi/2,\pi/2,\pi/2)$ spin-wave theory calculations become unstable 
(at $p_c \approx 0.5$) since the coefficient $A_{{\bf k}}$ becomes equal to $B_{{\bf k}}$. 
\begin{figure}[httb]
\centering
\includegraphics[width=3in]{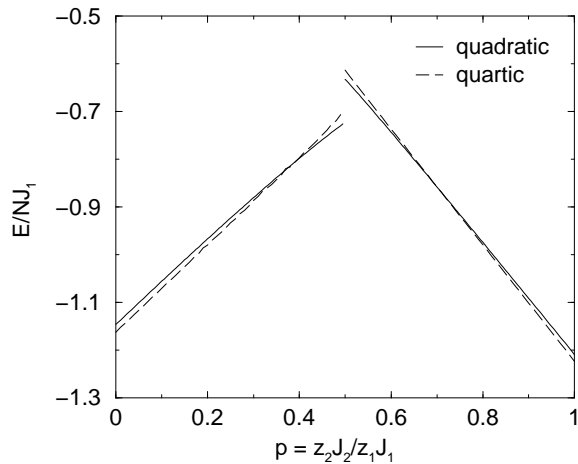}
\caption{\label{fig:Energy} Ground state energy per site, E/NJ$_1$, is plotted 
as a function of the frustration parameter $p=z_2J_2/z_1J_1$ without (solid 
lines) and with (dashed lines) quartic corrections for both AF$_1$ ($p<0.5$)
and AF$_2$ ($p>0.5$) ordered phases. For the bcc-lattice $z_1=8$ and $z_2=6$. 
Spin wave theory becomes unstable at the classical transition point, i.e. $p \approx 0.5$.
After extrapolation (not indicated in the figure above) we find that the two energies 
meet at $p \approx 0.53$ or $J_2/J_1 \approx 0.705$. The kink in the energy at this value 
of $p$ indicates a first order quantum phase transition from AF$_1$ to AF$_2$ phase.}
\end{figure}
In the AF$_2$ ordered phase or the lamellar phase with two interpenetrating N\'{e}el states, 
sublattice magnetization stays mostly flat except for a slight decrease (without quartic 
corrections) as $p$ approaches the critical value $p_c$ from above. The curve starts 
at $\approx 0.42$ for $p=1$ and ends at $\approx 0.41$ for $p=0.5$. However with quartic corrections
the curve has a very small upward turn. This upward curve has been observed in previous 
numerical works on this model.~\cite{oitmaa,Schmidt}
For the AF$_2$ phase, quartic fluctuations produce an overall enhancement of the magnetization over the 
high-$p$ values (0.5 to 1). But for low-$p$ (0 to 0.5), with increase in frustration 
quantum spin fluctuations play a dominant role as seen in Fig.~\ref{fig:submag}.

In Fig.~\ref{fig:Energy} we plot the ground state energy per site, E/NJ$_1$, 
for the AF$_1$ and AF$_2$ phases with and without quartic corrections
as a function of the frustration parameter $p=z_2J_2/z_1J_1$. $p_c=0.5$ is the 
classical transition point where a phase transition from the AF$_1$ phase to the 
AF$_2$ phase occur. The quadratic calculation agrees well with the results 
of ~\onlinecite{oitmaa,Schmidt}. The quartic corrections to the energy are shown by the 
dashed lines in Fig.~\ref{fig:Energy}. At $p=0$ calculated energy with the quartic correction
is slightly lower than the energy calculated without the quartic interaction terms. This 
small decrease from the linear spin wave theory calculation is due to the ground state energy
correction which is negative (as seen in Eq.~\ref{EquarticsmallJ}) from the quartic 
terms (self-energy Hartree diagrams). This trend for low-$p$ continues till 
$p \approx 0.38$ after which the energy with quartic corrections become dominant. 
For large $p$, we find the energy with quartic corrections to be lower than the energy 
calculated without the quartic interactions in the interval 
$\approx 0.70 - 1$. In both the phases quantum spin fluctuations tend to maintain the 
magnetic order by lowering the ground state energies. As $p$ approaches the critical value
$p_c$ from both phases, frustration increases causing the ground state energies to increase.
Then $1/S$ corrections due to spin fluctuations play a lesser role. 
As mentioned in the magnetization calculation 
our non-linear spin wave analysis becomes unstable at the classical transition 
point $p_c=0.5$. After extrapolation of the ground state energy curve from the 
AF$_1$ phase in the regime where non-linear spin wave theory breaks down we 
find that the energies from the two phases meet at $p \approx 0.53$.~\cite{note1} 
The kink at this point signals a first-order phase transition occurs from AF$_1$ to AF$_2$ 
phase. 
\section{\label{sec:conclusions} Conclusions}
In this work  we have investigated the zero temperature $1/S$ corrections to the 
sublattice magnetization and ground state energy of a  
spin-$1/2$ Heisenberg frustrated antiferromagnet on a bcc lattice using the framework of
non-linear spin wave theory. We have found that $1/S$ corrections due to spin-wave
interactions cause noticeable changes to the sublattice magnetization for both the 
two sublattice N\'{e}el phase (small NNN interaction $J_2$) and the AF$_2$ phase or the lamellar 
phase (large $J_2$). As non-linear spin wave theory calculations become unstable close to the classical 
transition point we are unable to analyze the nature of phase transition using this 
method. We also confirm that up to quartic corrections the system undergoes a 
first-order phase transition as indicated by a kink in the energy calculation. 
\section{Acknowledgment}
One of us (K.M.) thanks O. Starykh for helpful discussions.
\bibliography{BCC}

\begin{thebibliography}{31}
\bibitem[{not()}]{note1}
\bibinfo{note}{To avoid confusion we have not shown the extrapolated line in
  Fig.~\ref{fig:Energy}.}
\end{thebibliography}
\end{document}